\documentclass[conference]{IEEEtran}

\usepackage{cite}
\usepackage{graphicx}
\usepackage{graphicx}
\usepackage{subfigure}
\usepackage{float}
\usepackage{algorithm}
\usepackage{algorithmic}
\usepackage{mathrsfs}
\usepackage{amsfonts}
\usepackage{array}
\usepackage{amssymb,amsmath}
\usepackage{longtable}
\usepackage{rotating}
\usepackage{multirow}
\usepackage{cite}
\usepackage{float}
\usepackage[marginal]{footmisc}
\usepackage{stfloats}

\def\BibTeX{{\rm B\kern-.05em{\sc i\kern-.025em b}\kern-.08em
    T\kern-.1667em\lower.7ex\hbox{E}\kern-.125emX}}

\usepackage{bm}

\begin{document}
%
\title{Statistical CSI Based Hybrid mmWave MIMO-NOMA with Max-Min Fairness}


\author{\IEEEauthorblockN{Jinle Zhu$^\dag$, Qiang Li$^\dag$, Hongyang Chen$^\ddag$, and H. Vincent Poor$^\S$}\\
\IEEEauthorblockA{$^\dag$ National Key Laboratory of Science and Technology on Communications\\
University of Electronic Science and Technology of China, Chengdu, China\\
$^\ddag$ Zhejiang Lab, Hangzhou, China\\
$^\S$ Department of Electrical Engineering, Princeton University, Princeton, NJ 08544, USA\\
E-mail: {liqiang@uestc.edu.cn}}}

\maketitle

\vspace{-2cm}

\begin{abstract}
Non-orthogonal multiple access (NOMA) and millimeter wave (mmWave) are two key enabling technologies
for the fifth-generation (5G) mobile networks and beyond.
In this paper, we consider mmWave NOMA systems with max-min fairness constraints.
On the one hand, existing beamforming designs aiming at maximizing the spectrum efficiency (SE) are unsuitable for the NOMA systems with fairness in this paper.
On the other hand, previous work on about mmWave NOMA mostly depends on full knowledge of channel state information (CSI) which is extremely difficult to obtain accurately in mmWave communication systems.
To address this problem, we propose a heuristic hybrid beamforming design based on the statistical CSI (SCSI) user grouping strategy.
An analog beamforming scheme is first proposed to integrate the whole cluster users to mitigate the inter-cluster interference in the first stage.
Then two digital beamforming designs are proposed to further suppress the interference based on SCSI.
One is the widely used zero forcing (ZF) approach and the other is derived from the signal-to-leakage-plus-noise ratio (SLNR) metric extended from orthogonal multiple access (OMA) systems.
The effective gains fed back from the users are used for the power allocation. We introduce the quadratic transform (QT) method and bisection approach to reformulate this complex problem so as to rend it solvable.
Simulation results show that our proposed algorithms outperform the previous algorithms in term of user fairness.
\end{abstract}


\section{Introduction}

Recently, the explosive traffic growth envisioned in future wireless networks have  triggered and attracted tremendous research interests in millimeter wave (mmWave) communications due to their large bandwidth and the potential multiple access schemes \cite{mm}.
In contrast to the conventional sub-6 GHz frequency band, the wavelengths in the mmWave frequency band are short, which facilitates the deployment of massive antennas in a compact space to provide significant array gains to combat the propagation loss in the mmWave spectrum.
However, the traditional fully digital structure requires one dedicated radio frequency (RF) chain for each antenna, which incurs prohibitively high hardware cost and tremendous energy consumption in massive multiple-input multiple-output (MIMO) systems.
Hybrid architectures have been proposed as a feasible and compromise solution to strike a balance between energy consumption, signal processing complexity, and system performance, where
a large number of antennas are connected with a limited number of RF chains, which can reduce the power consumption of RF chains while maintaining highly directional beamforming to provide array gain \cite{HBF1,HBF2}.

Meanwhile, non-orthogonal multiple access (NOMA) has been recognized as an effective technique to further increase the spectral efficiency and support more connectivity \cite{Nall}.
The principle of NOMA is to server multiple users at the same time/frequency/code resource block (RB) by differentiating them in the power domain.
The success of NOMA is based on the successive interference cancelation (SIC) technique according to their channel condition and resource allocation.
Motivated by its promising advantages, the application of NOMA to mmWave communications with hybrid beamforming structure is a valuable study topic which has been studied in considerable research works \cite{Kmeans,CH,ICC,TWC,fair1,fair2}.

In \cite{Kmeans}, a machine learning based user clustering technique for mmWave NOMA systems has been proposed, where the users are assumed to be physically clustered in rooms or halls.
With this assumption, a K-means based user grouping scheme is naturally proposed.
This paper has proved that the K-means based user grouping scheme can achieve better performance than the cluster-head scheme \cite{CH}.
However, the K-means user grouping scheme is not generalized applicable to the common occasions where users are not physically clustered.
\cite{ICC} has proposed an agglomerative nesting (AGNES) based user grouping scheme which has been shown to outperform the K-means user grouping scheme when users are randomly positioned.
\cite{TWC} extended the work in \cite{ICC} to derive an enhanced joint user grouping and beam selection procedure.

\cite{Kmeans,CH,ICC} consider spectrum efficiency (SE) or energy efficiency (EE) as the evaluation criterion.
Thus, the beam patterns are pointed to the strongest user in each cluster to achieve a better performance.
However, this kind of beamforming design does not suitable for NOMA with fairness.
For instance, for a $\alpha\rightarrow\infty$ constrained NOMA system, i.e., the max-min fairness, the weak users barely benefit from the beamforming pointed to the strongest user, which will result in a poor SE performance since the users are required to be absolutely fair in the system.
Many works have investigated the fairness problem of NOMA systems \cite{fair1,fair2}, while they either considered the sub-6 GHz frequency band or failed to include the influence of beam alignment to the fairness among the users.
Furthermore, most works about NOMA \cite{Kmeans,CH,ICC,TWC} are based on the assumption that the transceivers have the full knowledge of the instantaneous channel state information (ICSI).
However, the fast variation of the mmWave channels makes it difficult to estimate the channel accurately.
Moreover, due to the large number of users in the NOMA systems and the massive antenna arrays at the base station (BS), the channel estimation results in a high training overhead.
Fortunately, the fading channel statistics of mmWave channels, such as angle of department (AoD) and large scale fading coefficient, are wide-dense stationary (WSS) due to its scattering-dependency \cite{r1}.
The feature that AoDs remain invariant over several periods of covariance time arouses an approach naturally fitting to a hybrid structure which is widely used in orthogonal multiple access (OMA) systems \cite{r2}.
The analog beamforming is based on the slowly-varying second order channel statistics and the digital beamforming matrix is designed based on the ICSI.

In this paper, we consider the max-min fairness problem of the downlink mmWave MIMO NOMA system which consists of user grouping, hybrid beamforming design, and power allocation.
In contrast to the previous mmWave MIMO-NOMA works, we propose to design the user grouping strategy and hybrid beamforming matrices based on the second order statistical CSI (SCSI) and the power allocation is based on the feedback of the effective channel gains which are scalars resulting in small feedback cost.
More specifically, a modified AGNES user grouping scheme based on the correlation matrix distance (CMD)is proposed.
At the analog beamforming stage, a concept of average channel covariance matrix is proposed since we regard all users in the cluster equally rather than focusing on the strongest user as in \cite{ICC,TWC}.
Two digital beamforming schemes are proposed to further suppress the inter-cluster interference.
One is to find the null space of the effective average covariance channel of each cluster and the other is inspired by the signal-to-leakage-plus-noise ratio (SLNR) metric which is widely used in OMA systems \cite{r3}.
We extend the SLNR metric in the NOMA system and obtain the corresponding digital beamforming vectors.
The power allocation is formulated as a max-min problem which can be solved by bi-section method and convex theory.
The simulation results suggest that under the proposed analog beamforming, the SLNR metric based digital beamforming can achieve approximate the same performance as the ZF digital beamforming. Besides, under the max-min fairness constraint, the proposed algorithms are shown to outperform the cluster-head beamforming schemes, which means that the proposed algorithms can balance the fairness among the users better than the previous designs.

Throughout this paper, upper-case and lower-case boldface letters denote matrices and vectors, respectively;
$(\cdot)^T$ and $(\cdot)^\dag $ denote the transpose and the Hermitian transpose of a matrix or a vector. $\mathcal{S}$ denotes a set;
$|\cdot|$ denotes the absolute value of a scalar or the cardinality of a set; $\|\cdot\|_2$ denotes the Frobenius norm of a vector or a matrix.
diag($\cdot$) represents a diagonal matrix.
$\mathbb{C}^{I\times J}$ denotes the set of all ${M\times N}$ matrices with complex entries. $\mathbb{E}\{\cdot\}$ denotes the expectation operation.
$\textbf{I}_{N}$ denotes a $N$ order unit matrix.

\section{System model}

We consider a downlink mmWave MIMO-NOMA system model.
The base station (BS) with $N_t$ antennas and $G$ RF chains serves $K$ single-antenna users simultaneously.
To simplify the description, the $k$-th user is denoted as $U_k$.
After the users are grouped into $G$ groups, the $u$-th user in the $g$-th group is denoted as $U_{g,u}$.
Hybrid beamforming is used to provide beam gain as well as suppress the interference among the users.
The signals first go through the digital beamforming matrix $\textbf{W}=[\textbf{w}_1,\textbf{w}_2,...,\textbf{w}_G]\in \mathbb{C}^{G\times G}$ and then are precoded by an analog beamforming matrix $\textbf{F}=[\textbf{f}_1,\textbf{f}_2,...,\textbf{f}_G]\in \mathbb{C}^{N_t\times G}$.
The received signal of $U_{g,u}$ is given as
\begin{align}
\nonumber r_{g,u}\hspace{-0.2em}=\sum\limits_{v=1}^{|\mathcal{S}_g|}\sqrt{P_{g,v}}\textbf{h}_{g,u}\textbf{F}\textbf{w}_gx_{g,v}+
\sum\limits_{q\neq g}^{G} \sum\limits_{v=1}^{|\mathcal{S}_q|}\sqrt{P_{q,v}}\textbf{h}_{g,u}\textbf{F}\textbf{w}_qx_{q,v},
\end{align}
where $\textbf{h}_{g,u}$, $x_{g,u}$, and $P_{q,v}$ are the channel information, transmitted signal and the allocated power of $U_{g,u}$.

At mmWave frequencies, the channels tend to be sparse and highly directional. We consider a general ray-based directional model with limited paths \cite{HBF2} which is defined as
\begin{equation}
\textbf{h}_{g,u}\hspace{-0.25em} =\hspace{-0.25em} \sqrt{\frac{N_t}{L_{g,u}}}\sum\limits_{l=1}^{L_{g,u}}\alpha_{g,u,l}\textbf{a}_{\rm BS}^\dag(\theta_{g,u,l})\triangleq \sqrt{\frac{N_t}{L_{g,u}}} \bar{\bm\alpha}_{g,u}\bm\Delta_{g,u}\textbf{A}_{g,u},
\end{equation}
where $L_{g,u}$ is the number of multipath components (MPC) of $U_{g,u}$.
$\alpha_{g,u,l}\sim \mathcal{CN}(0,\rho_{g,u,l}^2)$ denotes the small-scale fading of the $l$-th MPC.
$\rho_{g,u,l}^2$ can be viewed as the average power of the $l$-th MPC normalized by the large scale loss.
We assume that the MPCs are independent and we have $\sum_{l=1}^{L_{g,u}}\rho_{g,u,l}^2=1$.
The normalized path gain vector $\bar{\bm\alpha}_{g,u} = [\bar{\alpha}_{g,u,1},\bar{\alpha}_{g,u,2},...,\bar{\alpha}_{g,u,L_{g,u}}]$, where $\bar{\alpha}_{g,u,l} = \frac{\alpha_{g,u,l}}{\rho_{g,u,l}}$.
$\bm\Delta_{g,u} = {\rm diag}[\rho_{g,u,1},\rho_{g,u,2},...,\rho_{g,u,L_{g,u}}]$ stores the normalized average path gains $\{\rho_{g,u,l}\}$.
The beam steering vectors of AoD $\{\theta_{g,u,l}\}$ of $U_{g,u}$ are stacking in $\textbf{A}_{g,u}$, i.e., $\textbf{A}_{g,u}=[\textbf{a}_{\rm BS}(\theta_{g,u,1}),\textbf{a}_{\rm BS}(\theta_{g,u,2}),...,\textbf{a}_{\rm BS}(\theta_{g,u,L_{g,u}})]^\dag$.
Under an assumption of a uniform linear array (ULA), $\textbf{a}_{\rm BS}(\theta)$ can be written as
\begin{equation}
\textbf{a}_{\rm BS}(\theta) = \frac{1}{\sqrt{N_t}}\left[1,e^{j{\frac{2\pi d}{\lambda}} \sin\theta },...,e^{j{\frac{(N_t-1)2\pi d}{\lambda}} \sin\theta }\right],
\end{equation}
where $\lambda$ is the wavelength and $d=\frac{\lambda}{2}$ is the antenna spacing.

Under the block fading assumption, the small-scale fading vector remains constant within the coherence block and varies independently across blocks.
However, the fading channel statistics (angular power spectrum), i.e., $\bm\Delta_{g,u}$ and $\textbf{A}_{g,u}$ are WSS to tens or hundreds of coherence block lengths.
Hence, this statistical channel information can be evaluated efficiently by channel estimation techniques such as \cite{CE} with negligible normalized cost.
The long-term channel covariance matrix of $U_{g,u}$ can be calculated as
\begin{equation}
\textbf{R}_{g,u} = \mathbb{E}\{\textbf{h}_{g,u}^\dag \textbf{h}_{g,u}\}=\frac{N_t}{L_{g,u}}\textbf{A}_{g,u}^\dag\bm\Delta_{g,u}^2\textbf{A}_{g,u}.
\end{equation}
In Section III, we elaborate that our proposed user grouping and beamforming design depends only on the long-term SCSI.
In the power allocation stage, each user only needs to feed back an equivalent channel gain vector, which significantly reduces the information exchange overhead.

In mmWave MIMO-NOMA, multiple users are served in one beam where different beams (groups) will occur inter-group interference.
The users in the same group successively cancel the interference in a specific order which is referred as intra-group interference.
The decoding order is an essential issue which depends on not only the channel gain but also the beam gain and power allocation.
The details of the decoding order will be discussed in Section III.
We assume that the decoding order has been decided that $U_{g,u}$ can decode the signal of $U_{g,r}$ only if $r\geq u$.
The signal-to-interference-plus-noise ratio (SINR) of $U_{g,u}$ to decode its own signal can be denoted as
\begin{equation}\label{sr}
{{\rm SINR}_{g,u}}={\left|\textbf{h}_{g,u}\textbf{d}_{g}\right|^2 P_{g,u}}/
{ \eta_{g,u}},
\end{equation}
where $\textbf{d}_g=\textbf{F}\textbf{w}_g$ denotes the equivalent precoding vector for $\mathcal{S}_g$.
$\eta_{g,u}=I_{g,u}^{\rm intra}+I_{g,u}^{\rm inter} + {\sigma ^2}$ denotes the interference of $U_{g,u}$, which affects the decoding of its own signal.
$I_{g,u}^{\rm intra}=\sum^{u-1}_{v=1}\left|\textbf{h}_{g,u}\textbf{d}_{g}\right|^2 P_{g,v}$ denotes the intra-group interference and $I_{g,u}^{\rm inter}=\sum^G_{q\neq g}\sum^{|\mathcal{S}_q|}_{v=1}\left|\textbf{h}_{g,u}\textbf{d}_{q}\right|^2 P_{q,v}$ denotes the inter-group interference.
The SINR for $U_{g,u}$ decoding the signal of $U_{g,r}$ can be calculated as
\begin{equation}
{{\rm SINR}_{g,u,r}}={\left|\textbf{h}_{g,u}\textbf{d}_{g}\right|^2 P_{g,r}}/{ \eta_{g,u,r}},
\end{equation}
where $\eta_{g,u,r}=I_{g,u,r}^{\rm intra}+I_{g,u,r}^{\rm inter} + {\sigma ^2}$, $I_{g,u,r}^{\rm intra}=\sum^{r-1}_{v=1}\left|\textbf{h}_{g,u}\textbf{d}_{g}\right|^2 P_{g,v}$ and $I_{g,u,r}^{\rm inter}=I_{g,u}^{\rm inter}$.
To successfully perform SIC, the SINR condition should be satisfied which is given by $SINR_{g,u,r}\geq SINR_{g,u}$.

\section{PROBLEM FORMULATION AND PROBLEM SOLUTION}

\subsection{User grouping based on the SCSI}

Due to the principle of NOMA, an appropriate user group strategy will lay a solid foundation for the system performance because the user grouping directly influences the beam gain and power allocation scheme.
In mmWave-NOMA systems, the directionality of the mmWave channels enables us to cluster the users with high channel correlation in the same group and the users with low channel correlation in different groups.
Since the second order SCSI can capture the directionality feature of the channels, we use the CMD to show the distance between two channel covariance matrices \cite{r3}. The distance between $U_i$ and $U_j$ is given by
\begin{equation}\label{co1}
C(U_i,U_j)=1-
\frac{{\rm Tr}(\textbf{R}_i^\dag\textbf{R}_j)}{\|\textbf{R}_i\|_2\|\textbf{R}_j\|_2}.
\end{equation}
$C(U_i,U_j)$ measures the channel similarity between $U_i$ and $U_j$ ranging from 0 (the users' channels are completely uncorrelated)to 1 (the users' channels are completely orthogonal).
As in our previous work \cite{ICC}, we choose the complete linkage method to derive the linkage between cluster. We suppose that $\mathcal{S}_{i,j}$ is the user group merged from $\mathcal{S}_{i}$ and $\mathcal{S}_{j}$, namely, $\mathcal{S}_{i,j}\triangleq\mathcal{S}_{i}\cup\mathcal{S}_{j}$. Let $\mathcal{S}_{q}$ be one of the remaining groups except for $\mathcal{S}_{i}$ and $\mathcal{S}_{j}$. The complete linkage between $\mathcal{S}_{i,j}$ and $\mathcal{S}_{q}$ is given by
\begin{equation}\label{co2}
C(\mathcal{S}_{i,j},\mathcal{S}_{q})=\max\{C(\mathcal{S}_{i},\mathcal{S}_{q}), C(\mathcal{S}_{j},\mathcal{S}_{q})\}.
\end{equation}
The SCSI based AGNES algorithm is summarized in \textbf{Algorithm 1}.
\begin{algorithm}[!htbp]
\caption{AGNES clustering user grouping algorithm}

\textbf{Inputs}: Group number $G$, user set $\mathcal{U}=\{U_1,U_2,...,U_K\}$, second order SCSI ${{\textbf{R}}_{k}}$, $k=1,..,K$;\\
\textbf{Outputs}: User grouping strategy
 ${\Pi}=\{\mathcal{S}_1,\mathcal{S}_2,...,\mathcal{S}_G\}$;\\
\textbf{Initialization}: Initial single user groups $\mathcal{S}_k=\{U_k\},k=1,2,...,K$, group number index $t=K$.

\begin{algorithmic}[1]

\STATE Calculate the correlation $C$ in $\mathcal{U}$ by (\ref{co1}) and (\ref{co2});

\REPEAT

\STATE Search for two groups with the maximal similarity by the complete linkage method;

\STATE Merge the groups with the maximal similarity;

\STATE $t\leftarrow t-1$

\UNTIL {$t=G$}

\end{algorithmic}
\end{algorithm}

\subsection{Hybrid beamforming based on the SCSI}

After finishing user grouping, we design the hybrid beamforming matrices to provide beam gain.
In most previous works \cite{Kmeans,CH,ICC}, the beams are designed to point to the cluster head of each group since those works aim to maximize the sum rate of the systems.
Thus, it is reasonable that the users with best channel condition can get the best power resources.
While in this paper, we consider the fairness problem among the users.
The cluster head beamforming strategy only cares the strongest user in each group, which will directly omit the weak users.
With a forcing power allocation scheme to keep the fairness among the users, the data rate of the system is dragged due to the low beam gain of the users far from the strongest user in the angle domain.
We consider the user fairness before the power allocation stage and propose a new hybrid beamforming design scheme based on the second order SCSI.
The hybrid beamforming is difficult to design jointly because the constant modules (CM) constraint of the elements of the analog beamforming. Thus, we design the two matrices separately as the same in many hybrid beamforming works.

\paragraph {Analog beamforming} The basic idea is to design the analog beamforming matrix to enhance the beam gain of the whole cluster and suppress the interference from other clusters.
We regard the users in the same group equal in priority and they can be treated as a virtual multi-antenna user for this group with the average channel covariance matrix given by
\begin{equation}
\bar{\textbf{R}}_g = \frac{1}{|\mathcal{S}_{g}|}\sum\nolimits_{u=1}^{|\mathcal{S}_{g}|}\textbf{R}_{g,u}.
\end{equation}
The eigenvalue decomposition (EVD) of $\bar{\textbf{R}}_g$ is denoted as
\begin{equation}
\bar{\textbf{R}}_g = \textbf{U}_g{\bm\Lambda}_g\textbf{U}_g^\dag,
\end{equation}
where $\textbf{U}_g = [\textbf{u}_g^{\max}~\textbf{U}_g^N]$ and $\textbf{u}_g^{\max}\in\mathbb{C}^{N_t\times1}$ presenting the largest eigenvectors of $\bar{\textbf{R}}_g$.
We define the interference matrix of $\mathcal{S}_{g}$ with the dominant eigenvector of all other clusters as
\begin{equation}
\Xi_g = \left[\textbf{u}_1^{\max},...,\textbf{u}_{g-1}^{\max},\textbf{u}_{g+1}^{\max},...,\textbf{u}_G^{\max}\right]
\in\mathbb{C}^{N_t\times(G-1)}.
\end{equation}
We perform the singular-value decomposition (SVD) of the interference matrix as
\begin{equation}
\Xi_g = \textbf{Q}_g\bm\Sigma_g\textbf{V}_g^\dag,
\end{equation}
where $\textbf{Q}_g = \left[\textbf{Q}_g^O~\textbf{Q}_g^N\right]$, with $\textbf{Q}_g^N\in\mathbb{C}^{N_t\times(N_t-(G-1))}$ representing the null space of $\Xi_g$.
By connecting the projection the space of $\textbf{Q}_g^N$ and the average channel covariance matrix of ${\mathcal{S}_{g}}$, we give the effective channel covariance $\hat{\textbf{R}}_g$ as
\begin{equation}
\hat{\textbf{R}}_g = \hat{\textbf{U}}_g\hat{{\bm\Lambda}}_g\hat{\textbf{U}}_g^\dag =
(\textbf{Q}_g^N)^\dag\textbf{U}_g{\bm\Lambda}_g\textbf{U}_g^\dag\textbf{Q}_g^N,
\end{equation}
where $\hat{\textbf{U}}_g = \left[\hat{\textbf{u}}_g^{\max}~\hat{\textbf{U}}_g^N\right]$, with $\hat{\textbf{u}}_g^{\max}\in\mathbb{C}^{(N_t-(G-1))\times1}$ representing the largest eigenvector of $\hat{\textbf{R}}_g$.
The analog beamforming vector corresponding to $\mathcal{S}_g$ can be calculated as
\begin{equation}
\textbf{f}_g = \textbf{Q}_g^N\hat{\textbf{u}}_g^{\max}.
\end{equation}

\paragraph {Digital beamforming}
Since the analog beamforming cannot fully cancel the inter-group interference. The digital beamforming is used to further mitigate the interference. After obtaining $\textbf{F}$, the BS can get the effective channel of $U_{g,u}$ as $\textbf{h}_{g,u,{\rm eff}} =\textbf{h}_{g,u}\textbf{F}\in\mathbb{C}^{1\times G}$, where the effective channel covariance matrix can be calculated as $\textbf{R}_{g,u,{\rm eff}}=\textbf{F}^\dag\textbf{R}_{g,u}\textbf{F}\in\mathbb{C}^{G\times G}$.
With the central covariance matrix $\bar{\textbf{R}}_g$, the effective average channel covariance matrix of $\mathcal{S}_{g}$ is denoted as $\bar{\textbf{R}}_{g,{\rm eff}}=\textbf{F}^\dag\bar{\textbf{R}}_{g,u}\textbf{F}\in\mathbb{C}^{G\times G}$.
The digital beamforming vector of $\mathcal{S}_{g}$ can be designed in the nullspace of the dominant eigenvectors of the effective channel matrices of the other clusters for $q\neq g$, i.e.,
\begin{equation}
\textbf{w}_g^\dag\textbf{u}_{\max}(\bar{\textbf{R}}_{q,{\rm eff}}) =
\begin{cases}
0& {g\neq q},\\
1& {g=q},
\end{cases}
\end{equation}
where $\textbf{u}_{\max}(\bar{\textbf{R}}_{q,{\rm eff}})$ denotes the largest eigenvector of $\bar{\textbf{R}}_{q,{\rm eff}}$.
Thus, with zero forcing (ZF) method, the digital beamforming matrix can be obtained by
\begin{equation}\label{e15}
\textbf{W} = \textbf{U}_{\max}^\dag \left(\textbf{U}_{\max}\textbf{U}_{\max}^\dag\right)^{-1},
\end{equation}
where $\textbf{U}_{\max} = \left[\textbf{u}_{\max}(\bar{\textbf{R}}_{1,{\rm eff}}),\textbf{u}_{\max}(\bar{\textbf{R}}_{2,{\rm eff}}),\cdots,\textbf{u}_{\max}(\bar{\textbf{R}}_{G,{\rm eff}})\right]$ collects the eigenvectors of the effective average channel covariance matrices.
This design intends to reduce the inter-group interference.
Another perspective is to maximize the desired power and suppress the leakage power to other clusters.
Inspired by this, we derive the SLNR metric to strike a balance between the desired signal power and the interference.
In OMA systems, SLNR is a widely used metric where each digital beamforming vector is for one single user \cite{r3}.
However, each digital beamforming vector is corresponding to a group with multiple users. Thus, we define SLNR of $\mathcal{S}_{g}$ as
\begin{equation}
{\rm SLNR}_g=\frac{\sum\limits_{v=1}^{|\mathcal{S}_{g}|}P_{g,v}\left|\textbf{h}_{g,v,{\rm eff}}^\dag \textbf{w}_g\right|^2}{\sigma^2+P_{g}\sum\limits_{q\neq g}^G\sum\limits_{v=1}^{|\mathcal{S}_{q}|}
\left|\textbf{h}_{q,v,{\rm eff}}^\dag \textbf{w}_g\right|^2},
\end{equation}
where $P_g=\sum_{v=1}^{|\mathcal{S}_{g}|}P_{g,v}$ denotes the power allocated to $\mathcal{S}_{g}$.
The numerator of ${\rm SLNR}_g$ is the combination gain consists of the channel gain and beamforming gain of $\mathcal{S}_{g}$.
The denominator is the leakage power to other clusters for $q\neq g$.
The lower bound on the average SLNR can be derived as
\begin{align}\label{SLNR}
\nonumber&\mathbb{E}({\rm SLNR}_g) = \\
\nonumber &\mathbb{E}({\sum\limits_{v=1}^{|\mathcal{S}_{g}|}P_{g,v}\left|\textbf{h}_{g,v,{\rm eff}}^\dag \textbf{w}_g\right|^2})\mathbb{E}(\frac{1}{\sigma^2+P_{g}\sum\limits_{q\neq g}^G\sum\limits_{v=1}^{|\mathcal{S}_{q}|}
\left|\textbf{h}_{q,v,{\rm eff}}^\dag \textbf{w}_g\right|^2})\\
&\geq \frac{\textbf{w}_g^\dag\left(\sum\limits_{v=1}^{|\mathcal{S}_{g}|}P_{g,v}\textbf{R}_{g,v,{\rm eff}}\right)\textbf{w}_g}{\sigma^2+
P_{g}\textbf{w}_g^\dag\left(\sum\limits_{q\neq g}^G\sum\limits_{v=1}^{|\mathcal{S}_{q}|}
\textbf{R}_{q,v,{\rm eff}}\right)\textbf{w}_g}\triangleq {\rm SLNR}_g^{\rm{LB}}.
\end{align}
The first equation in (\ref{SLNR}) is from the dependence between the numerator and the denominator and the inequation is obtained by the convexity of $f(x)=1/x$.
\cite{SLNR} has proved that the optimal unit norm $\textbf{w}_g$ to maximize ${\rm SLNR}_g^{\rm{LB}}$ is the generalized eigenvector given by
\begin{equation}\label{e18}
\textbf{w}_g = \textbf{u}_{\max}\left(
\frac{\sum\limits_{v=1}^{|\mathcal{S}_{g}|}P_{g,v}\textbf{R}_{g,v,{\rm eff}}}{\sigma^2\textbf{I}_{G}+
P_{g}\sum\limits_{q\neq g}^G\sum\limits_{v=1}^{|\mathcal{S}_{q}|}
\textbf{R}_{q,v,{\rm eff}}}\right).
\end{equation}
Due to the beamforming power constraint, the digital beamforming vector can be normalized as $\textbf{W}=\frac{\sqrt{G}\textbf{W}}{\|\textbf{FW}\|_2}$.

\subsection{Max-Min Fairness Power Allocation}

In Subsection B, we design the beamforming matrices to consider the other weak users rather than straightforwardly pointing the beam to the strongest user, so that the fairness can be imposed in the beamforming aspect.
After obtaining the beamforming matrices, we reorder the users in the same group to decide the decoding order based on the effective channel gain feedback.
Each user feeds back the effective channel gain vectors, i.e., $\left[\left|\textbf{h}_{g,u}\textbf{d}_{1}\right|^2,\left|\textbf{h}_{g,u}\textbf{d}_{2}\right|^2,\cdots, \left|\textbf{h}_{g,u}\textbf{d}_{G}\right|^2\right]$, according to the fixed beamforming matrices.
The users are ordered as $\left|\textbf{h}_{g,i}\textbf{d}_{g}\right|^2\geq\left|\textbf{h}_{g,j}\textbf{d}_{g}\right|^2$ in $\mathcal{S}_g$ for $i<j$, $g=1,\cdots,G$.
To ensure user fairness, proper power allocation is essential.
Consider the fairness of the other weak users in the system, we choose max-min fairness as the criterion which aims to maximize the minimum achievable user rate in the cluster.
The problem can be formulated as follows
\begin{align}
\textbf{P1}~~~~&\mathop {\max}\limits_{\left\{ P_{g,u} \right\}} \mathop{\min}\limits_{\{g,u\}} Rate_{g,u} \label{eq:22}\\
&{\rm{s.t.}}~{\rm{C}}1:\sum\limits_{g = 1}^G \sum\limits_{u = 1}^{\left| {{\mathcal{S}_g}} \right|}{ {{P_{g,u}}}}  \le {P_{\max }},\tag{\ref{eq:22}$\mathrm{a}$}\\
&~~~~~{\rm{C}}2:{R_{g,u}} \ge {R_{\min}},\tag{\ref{eq:22}$\mathrm{b}$}\\
\nonumber&~~~~~~~~~~~\forall g=1,2,...,G, u=1,2,...,{\left| {{\mathcal{S}_g}} \right|},\\
&~~~~~{\rm{C}}3:SINR_{g,u,r}\geq SINR_{g,r},\tag{\ref{eq:22}$\mathrm{c}$}\\
\nonumber&~~~~~~~~~~~\forall g=1,2,...,G, r=u+1,...,{\left| {{\mathcal{S}_g}} \right|},
\end{align}
where $Rate_{g,u}=\log2(1+{\rm SINR}_{g,u})$ is the achievable data rate of $U_{g,u}$.
C1 is the total power constraint for the BS.
C2 is the QoS guarantee for each user.
With the given decoding order, C3 is a constraint to ensure the SIC can be performed successfully.
$\textbf{P1}$ is a non-convex problem because its objective function is non-convex.
We introduce a variable $t$ to denote the minimum achievable rate of the users.
Thus, $\textbf{P1}$ can be re-written as $\textbf{P2}$.
\begin{align}
\textbf{P2}~~~~&\mathop {\max}\limits_{\left\{ P_{g,u} \right\},t} t \label{eq:P2}\\
&\nonumber {\rm s.t}.~~{\rm{ C1,C3,}}\\
& \nonumber{\rm{C}}4: R_{g,u}\geq\max\{R_{\min},t\}.\tag{\ref{eq:P2}$\mathrm{a}$}
\end{align}
C4 is the combination constraint of $R_{g,u}>t$ and C2.
In \textbf{P2}, ${\left\{ P_{g,u} \right\}}$ and $t$ are entangled, which makes the problem difficult to solve directly.
However, the introduced variable $t$ is a one-dimension scalar. With a fixed range, $t$ can be searched by the bisection method.
As $t$ presents the minimum achievable rate of all users, the lower bound $t_{\min}$ should be the $R_{\min}$ under the QoS constraint.
We consider an extreme case that the BS generates one beam with all power pointing to the strongest user among the $K$ users.
Assuming $U_i$ is the strongest user, the beam pattern can be obtained by the equal gain transmission (EGT) algorithm in \cite{EGT} which can be calculated by $\textbf{f}_\star = \measuredangle\textbf{u}_{\max}^\dag\left(\textbf{R}_i\right)/\sqrt{(N_t)}$, where $\measuredangle$ means the angle extraction operation.
The initial upper bound can be calculated by $t_{\max} = \frac{\left|\textbf{f}_\star^\dag\textbf{R}_i \textbf{f}_\star\right|}{N_t\sigma^2}$.
With a specific constant value $t$, we evaluate that whether there exists a feasible power solution by solve the following problem
\begin{align}
\textbf{P3}~~~~&\mathop {\min}\limits_{\left\{ P_{g,u} \right\}} \sum\limits_{g = 1}^G \sum\limits_{u = 1}^{\left| {{\mathcal{S}_g}} \right|}{ {{P_{g,u}}}} \label{eq:P3}\\
&\nonumber {\rm s.t}.~~{\rm{C3}},\\
& \nonumber{\rm{C}}4: 2y_{g,u}\sqrt{\left|\textbf{h}_{g,u}\textbf{d}_{g}\right|^2P_{g,u}}-y_{g,u}^2\eta_{g,u}\geq 2^\varrho-1,\tag{\ref{eq:P3}$\mathrm{a}$}
\end{align}
where $\varrho=\max\{R_{\min},t\}$ and $\{y_{g,u}\}$ is the auxiliary variable in the quadratic transform \cite{QT}.
With the fixed $t$ and $y_{g,u}$, $\textbf{P}3$ is a convex problem as to $P_{g,u}$.
$y_{g,u}$ can be updated by
\begin{equation}\label{SEm}
y_{g,u}^{\star} = \sqrt{\left|\textbf{h}_{g,u}\textbf{d}_{g}\right|^2P_{g,u}}/
\eta_{g,u}.
\end{equation}
If the sum power {\small $\sum\nolimits_{g = 1}^G \sum\nolimits_{u = 1}^{| {{\mathcal{S}_g}} |}{ {{P_{g,u}}}} \leq P_{\max}$}, the power satisfies the minimum data rate constraint as well as the sum power limit, which implies that the mininum rate bound $t_{\min}$ should be larger than the current $t$.
Otherwise, the maximum achievable data rate $t_{\max}$ should be smaller than the current $t$. By appropriately adjusting the bounds of $t$ by the bisection procedure, $t$ and $\{P_{g,u}\}$ can be found within a desirable accuracy $\tau$.
It should be noted that $\left(\ref{eq:P2}\mathrm{a}\right)$ and $\left(\ref{eq:P3}\mathrm{a}\right)$ are two equivalent constraints.
The equivalence and the convexity of $\textbf{P}3$ can be found in \cite{QT}.
The whole power allocation algorithm is summarized in \textbf{Algorithm2}.

\begin{algorithm}
\caption{Power Allocation}
\textbf{Inputs:} User grouping strategy $\Pi$, $\textbf{F}_{BB}$, $\textbf{F}_{RF}$, $\textbf{R}_{g,u}$.\\
\textbf{Outputs:} Power Allocation $\{P_{g,u}\}$;

\begin{algorithmic}[1]

\REPEAT

\STATE $t = {t_{\min}+t_{\max}}/{2}$ ;

\REPEAT

\STATE Update $y_{g,u}^\star$ by (\ref{SEm});

\STATE Update $P_{g,u}$ by solving the convex optimization problem $\textbf{P3}$ for fixed $\textbf{y}$ and $t$;

\UNTIL {$\textbf{P}$ converges.}

\IF {$\sum\nolimits_{g = 1}^G \sum\nolimits_{u = 1}^{\left| {{\mathcal{S}_g}} \right|}{ {{P_{g,u}}}} \leq P_{\max}$}

\STATE $t_{\min} = t$;

\ELSE

\STATE $t_{\max} = t$;

\ENDIF

\UNTIL {$t_{\max}-t_{\min}<\tau$.}

\end{algorithmic}
\end{algorithm}

\section{SIMULATION RESULTS}

In this section, we evaluate the performance of the proposed SCSI based mmWave MIMO-NOMA through Monte Carlo simulations.
Different user grouping schemes and beamforming designs are tested and compared.
The BS is amounted with $N_t=64$ transmitting antennas connected to $G=4$ RF chains.
Without loss of generality, $L_{g,u}=6$ component paths are assumed for each user.
The DoA parameters $\theta_{g,u,l}$ are uniformly distributed within $[-\frac{\pi}{2},\frac{\pi}{2}]$.
The accuracy for power allocation convergence is set as $\tau=10^{-5}$.
The QoS minimum rate constraint for each user is $R_{\min}=0.01 {\rm bps/Hz}$ \cite{Nall}.
We set ${\rm SNR}=\frac{E_b}{\sigma^2}$.
\begin{figure}[hbpt]
	\centering
	\subfigure[Minimum data rate vs. $P_{\max}$.]{
		\label{Pmax}
		\includegraphics[width=0.22\textwidth]{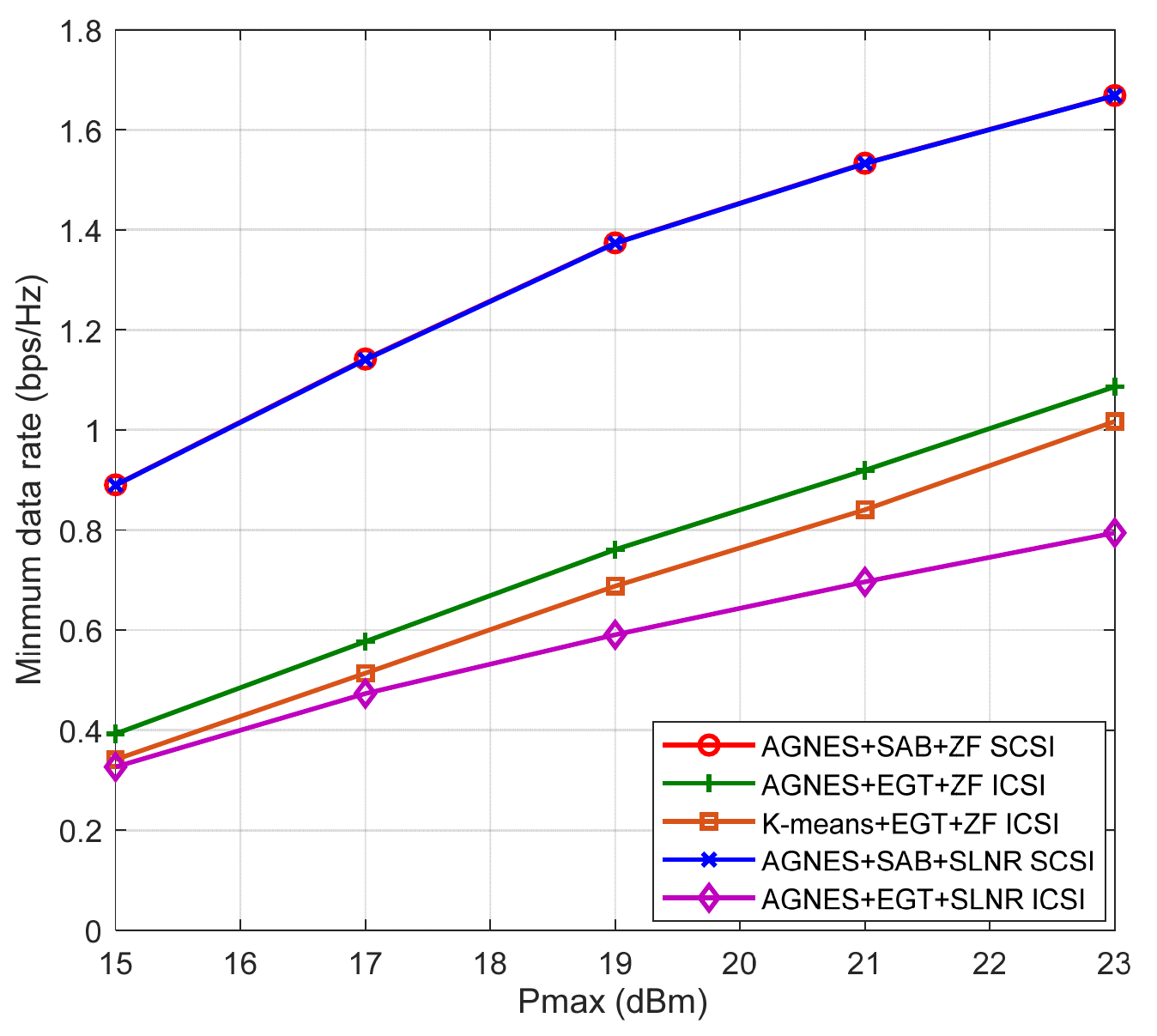}}
	\subfigure[Sum rate MMSE vs. $K$.]{
		\label{K}
		\includegraphics[width=0.22\textwidth]{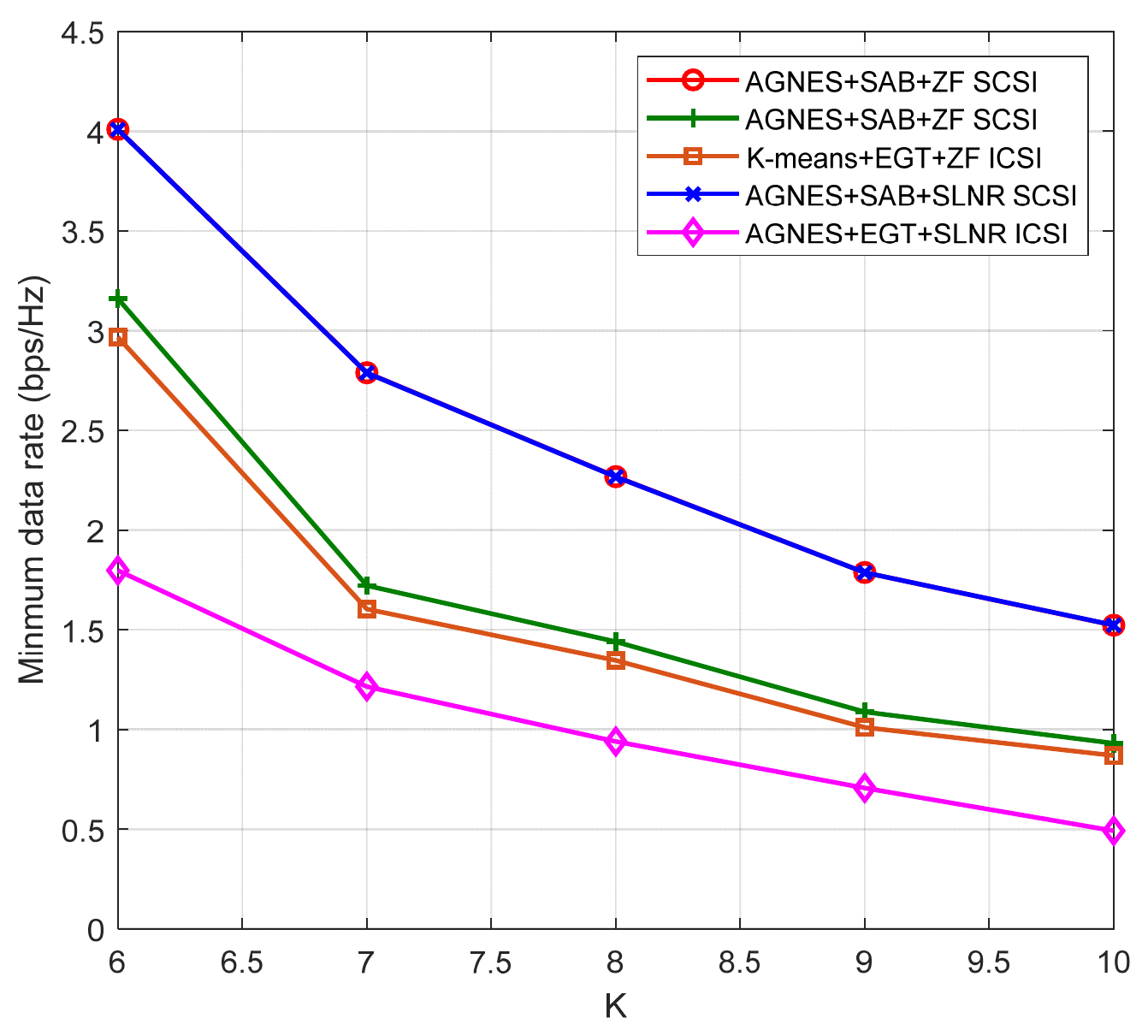}}
	\caption{Performance comparison.  }
	\label{fig:1}
\end{figure}

Next, we briefly introduce the schemes compared in the simulation.
For the user grouping phase, there are two schemes, i.e., AGNES and K-means, where AGNES can be applied based on ICSI \cite{ICC} or SCSI and K-means is used in the scheme with ICSI.
The proposed analog beamforming matrix design is denoted as ``Statistical analog beamforming (SAB)".
We also proposed two digital beamforming designs, where one by (\ref{e15}) is based on ZF method and the other one with SLNR metric by (\ref{e18}) can be applied based on either ICSI or SCSI.
In the classical cluster-headed scheme, the analog beamforming vector is chosen from the codebook to maximize the channel gain \cite{Kmeans} and the digital beamforming is designed by choosing the strongest effective channel and perform ZF.
To fairly compare the schemes, we calculate the analog beamforming vectors with EGT to harvest the large array gain.
We match several schemes to better observe the influence of each stage.
For instance, ``AGNES+EGT+SLNR ICSI" represents that we choose AGNES user grouping scheme, EGT analog beamforming scheme and SLNR digital beamforming scheme based on ICSI.


In Fig. \ref{Pmax}, we show the minimum spectrum efficiency performance of the different algorithms with the settings to  be $K=9$ and ${\rm SNR}=0~{\rm dB}$.
We see that with the proposed statistical analog beamforming scheme SAB, the ZF method and the SLNR based metric can achieve approximate performance, which means that either eliminating the effective inter-cluster interference or suppress the cluster leakage power can produce good effect.
However, this only happens when proposed analog beamforming scheme is used.
When only the strongest users in each cluster are chosen, i.e., the analog beamforming is calculated based on EGT, the algorithm based on SLNR is less effective than the ZF algorithm.
The reason is that the EGT algorithm only takes the strong users into account which omits the interference to the weak users.
The SLNR metric fails to decrease the leak interference due to this omission.
When ICSI is assumed, the AGNES based user grouping scheme is better than the K-means user grouping scheme because the AGNES algorithm can form the cluster spontaneously and does not depend on the initial points used by the K-means algorithm.
This is consistent with the conclusion in \cite{ICC}.


In Fig. \ref{K}, the performance of the minimum data rate versus the number of users of different algorithms is shown.
We set ${\rm SNR}=5 ~{\rm dB}$ and ${\rm Pmax} = 21~{\rm dBm}$.
We see that the minimum data rate decreases when the number of users $K$ increases for all algorithms since the average power for each user decreases.
Under SCSI, the SLNR based digital beamforming design also show good performance to be approximately the same as the ZF digital beamforming design.
This means that the SLNR metric can successfully suppress the inter-cluster interference when considering the interference leakage metric even in the overload mmWave MIMO-NOMA system.

\section{CONCLUSION}

In this paper, we have studied the fairness problem of mmWave MIMO-NOMA systems.
We have consider a more practical scenario that only second order SCSI has been obtained in the user grouping and hybrid beamforming phases.
A user grouping scheme based on the SCSI and AGNES algorithm has been proposed which allows more than one user in a cluster.
Hybrid beamforming schemes have been provided to promote the fairness among the users in each cluster and reduce the inter-cluster interference.
Power allocation with QT has been solved using convexity tools to ensure the fairness for NOMA.
The simulation results show that the proposed beam alignment balances the beam gain in a cluster better, which balances the SE performance with fairness.

\end{document}